\documentclass[12pt]{article}
\begin{document}
\title{GRAVITATION REVISITED}
\author{B.G. Sidharth\\
International Institute for Applicable Mathematics \& Information Sciences\\
Hyderabad (India) \& Udine (Italy)\\
B.M. Birla Science Centre, Adarsh Nagar, Hyderabad - 500 063 (India)}
\date{}
\maketitle
\begin{abstract}
Gravitation has posed a puzzle and a problem for many decades. Attempts to unify it with other fundamental interactions have failed. These problems and puzzles have been underscored by the likes of Witten and Weinberg. We survey this and argue that gravitation has a different character compared to other fundamental interactions - it is an energy distributed over all the elementary particles in the universe. The above puzzle and problem is resolved satisfactorily. These considerations lead to a varying $G$ cosmology consistent with observation. It is argued that apart from the usual tests, the above explains in addition the anomalous accelerations of the Pioneer spacecrafts. Further tests are also proposed.
\end{abstract}
\section{Introduction}
Gravitation has been a puzzle and a challenge for several decades. As Witten put it \cite{witten}, ``The existence of gravity clashes with our description of the rest of physics by quantum fields''. Indeed after Einstein's formulation of gravitation, a problem that has defied solution has been that of providing a unified description of gravitation along with other fundamental interactions. One of the earliest attempts was that of Hermann Weyl - the gauge geometry \cite{weyl}, which though elegant was rejected on the grounds that in the final analysis, it was not really a unification of gravitation with electromagnetism.\\
Modern approaches to this problem have lead to the abandonment of a smooth spacetime manifold. Instead, the Planck scale is now taken to be a minimum fundamental scale. In particular, the author's 1997 work, to be touched upon, threw up a dark energy driven accelerating universe with a small cosmological constant. In this model, the gravitational constant $G$ varies with time as in the Dirac cosmology. \\
Observations from 1998 onwards have shown that the universe is indeed accelerating with a small cosmological constant and that dark energy, rather than dark matter predominates \cite{kirsh}.\\ 
Cosmologies with time varying $G$ have also been considered in the past, for example in the Brans- Dicke theory or in the Dirac large number theory or by Hoyle \cite{narlikar,barrow,nar2,nar3,cu}. In the case of the Dirac cosmology, the motivation was Dirac's observation that the supposedly large number coincidences involving $N$, the number of elementary particles in the universe had an underlying message if it is recognized that
\begin{equation}
\sqrt{N} \propto T\label{e1}
\end{equation}
where $T$ is the age of the universe. Equation (\ref{e1}) lead to a $G$ decreasing inversely with time.\\
The Brans-Dicke cosmology arose from the work of Jordan who was motivated by Dirac's ideas to try and modify General Relativity suitably. In this scheme the variation of $G$ could be obtained from a scalar field $\phi$ which would satisfy a conservation law. This scalar tensor gravity theory was further developed by Brans and Dicke, in which $G$ was inversely proportional to the variable field $\phi$. (It may be mentioned that more recently the ideas of Brans and Dicke have been further generalized.)\\
In the Hoyle-Narlikar steady state model, it was assumed that in the Machian sense the inertia of a particle originates from the rest of the matter present in the universe. This again leads to a variable $G$. The above references give further details of these various schemes and their shortcomings which have lead to their fall out of favour.\\
In the author's 1997 cosmology particles were fluctuationally created from a background dark energy, due ultimately to Planck scale effects, in an inflationary type phase transition and this lead to a scenario of an accelerating universe with a small cosmological constant, which was observationally confirmed in 1998 itself by Perlmutter, Schmidt and coworkers as also by the Wilkinson Microwave Anisotropy Probe and the Sloan Digital Sky Survey in 2003. The details of this paradigmatic shift in cosmology are all discussed in references \cite{ijmpa,ijtp,csfcos,csfcos2,perl,science, science2,cu,uof}. Moreover in the author's cosmology the various supposedly miraculous large number coincidences as also the otherwise inexplicable Weinberg formula which gives the mass of an elementary particle in terms of the gravitational constant and the Hubble constant are also deduced from the underlying theory rather than being ad hoc \cite{weinberg}. (In fact Weinberg noted that his pion mass-Hubble constant relationship cannot be dismissed as accidental but rather needed to be explained. We will come back to this later).  The gravitational constant is given again by 
\begin{equation}
G = \frac{G_0}{T}\label{e2}
\end{equation}
where $T$ is the time (the age of the universe) and $G_0$ is a constant. Further other routine effects like the precession of the perihelion of Mercury and the bending of light, the flattening of rotational curves of galaxies etc. are also explained in this model \cite{nc115,cu}. Moreover in this model, $\Lambda$ is given by $\Lambda < 0 (H^2)$ and shows the inverse dependence $1/T^2$ on time. Already there is observational evidence for (\ref{e2}) as will be noted in section 6.\\
With this background, first we now give some further tests for equation (\ref{e2}).
\section{The Anomalous Acceleration of the Pioneer Spacecrafts}
The inexplicable anomalous accelerations of the Pioneer spacecrafts which have been observed by J.D. Anderson and coworkers at the Jet Propulsion Laboratory for well over a decade \cite{csfkerr,anderson} have posed a puzzle. This can be explained by (\ref{e2}). This can be seen in a simple way as follows: Infact from the usual orbital equations we have \cite{gold}
$$v \dot {v} \approx -\frac{GM}{2tr} (1 + e cos \Theta )-\frac{GM}{r^2} \dot {r}(1+e cos \Theta )$$
$v$ being the velocity of the spacecraft. It must be observed that the first term on the right side is the new effect due to (\ref{e2}). There is now an anomalous acceleration given by
$$a_r = \langle \dot v \rangle_{\mbox{anom}} = \frac{-GM}{2t r v} (1+e cos \Theta )$$
$$\approx -\frac{GM}{2t\lambda} (1+e)^3$$
where 
$$\lambda = r^4 \dot \Theta^2$$
If we insert the values for the Pioneer spacecrafts we get 
$$a_r \sim -10^{-7} cm/sec^2$$
This is the anomalous acceleration reported by Anderson and co-workers.\\
We will next deduce that the equation (\ref{e2}) also explains correctly the observed decrease in the orbital period of the binary pulsar $PSR\, 1913 + 16$, which has also been attributed to as yet undetected gravitational waves \cite{ohan}.
\section{The Binary Pulsar}
It may be observed that the energy $E$ of two masses $M$ and $m$ in gravitational interaction at a distance $L$ is given by
\begin{equation}
E = \frac{GMm}{L} = \mbox{constant}\label{e3}
\end{equation}
We note that if this energy decreases by any mechanism, for example by the emission of gravitational waves, or by the decrease of $G$, then because of (\ref{e3}), there is a compensation by the decrease in the orbital length and orbital period. This is the standard General Relativistic explanation for the binary pulsar $PSR\, 1913 + 16$. We will show that the same holds good, if we are given (\ref{e2}). In this case we have, from (\ref{e3})
\begin{equation}
\frac{\mu}{L} \equiv \frac{GMm}{L} = \mbox{const.}\label{e4}
\end{equation}
Using (\ref{e2}) we can write, for a time increase $t$,
\begin{equation}
\mu = \mu_0 - Kt\label{e5}
\end{equation}
where we have
\begin{equation}
K \equiv \dot{\mu}\label{e6}
\end{equation}
In (\ref{e6}) $\dot{\mu}$ can be taken to be a constant in view of the fact that $G$ varies very slowly with time as can be seen from (\ref{e2}). Specifically we have
\begin{equation}
G (T + t) = G(T) -t \frac{G}{T} + \frac{t^2}{2} \frac{G}{T^2} + \cdots \approx G(T) - t \frac{G}{T}\label{e7}
\end{equation}
where $T$ is the age of the universe and $t$ is an incremental time. Whence using (\ref{e7}) $K$ in (\ref{e6}) is given by
$$K \propto \frac{G}{T}$$
and so 
$$\dot{K} \sim \frac{G}{T^2} \approx 0$$
So (\ref{e4}) requires
$$L = L_0 (1 - \alpha K)$$
Whence on using (\ref{e5}) we get
\begin{equation}
\alpha = \frac{t}{\mu_0}\label{e8}
\end{equation}
Let us now consider $t$, to be the period of revolution in the case of the binary pulsar. Using (\ref{e8}) it follows that
\begin{equation}
\delta L = - \frac{L_0tK}{\mu_0}\label{e9}
\end{equation}
We also know (Cf.ref.\cite{gold})
\begin{equation}
t = \frac{2\pi}{h} L^2 = \frac{2\pi}{\sqrt{\mu}}\label{e10}
\end{equation}
\begin{equation}
t^2 = \frac{4\pi^2 L^3}{\mu}\label{e11}
\end{equation}
Using (\ref{e9}), (\ref{e10}) and (\ref{e11}), a little manipulation gives
\begin{equation}
\delta t = - \frac{2t^2K}{\mu_0}\label{e12}
\end{equation}
(\ref{e9}) and (\ref{e12}) show that there is a decrease in the size of the orbit, as also in the orbital period. Such a decrease in the orbital period has been observed in the case of binary pulsars \cite{ohan,will}.\\
Let us now apply the above considerations to the case of the binary pulsar $PSR\, 1913 + 16$ observed by Taylor and coworkers (Cf.ref.\cite{will}). In this case it is known that, $t$ is 8 hours while $v$, the orbital speed is $3 \times 10^7 cms$ per second. It is easy to calculate from the above
$$\mu_0 = 10^4 \times v^3 \sim 10^{26}$$
which gives $M \sim 10^{33}gms$, which of course agrees with observation. Further we get using (\ref{e1}) and (\ref{e5})
\begin{equation}
\Delta t = \eta \times 10^{-5} sec/yr, \eta <\approx 8\label{e13}
\end{equation}
Indeed (\ref{e13}) is in good agreement with the carefully observed value of $\eta \approx 7.5$ (Cf.refs.\cite{ohan,will}).\\
It should also  be remarked that in the case of gravitational radiation, there are some objections relevant to the calculation (Cf.ref.\cite{will}).
\section{Change in Orbital Parameters}
To consider the above result in a more general context, we come back to the well known orbital equation \cite{gold}
\begin{equation}
d^2 u/d\Theta^2 + u = \mu_0/h^2\label{e14}
\end{equation}
where $\mu_0 = GM$\\
$M$ is themass of the central object and $h - r^2 d\Theta / dt$ - a constant and $u - \frac{1}{r}$. The solution of (\ref{e14}) is well known,
$$lu = 1 + ecos \Theta$$
where $l = h^2/\mu_0$.\\
It must be mentioned that in the above purely classical analysis, there is no precession of the perihelion.\\
We now replace $\mu_0$ by $\mu$ and also assume $\mu$ to be varying slowly as $G$ varies slowly and uniformly as earlier:
\begin{equation}
\dot{\mu} = d\mu / dt = k, \mbox{a \, constant}\label{e15}
\end{equation}
as $\dot{k} \sim 0 \frac{1}{T^2}$ and can be neglected.\\
Using (\ref{e15}) in (\ref{e14}) and solving the orbital equation (\ref{e14}), the solution can now be obtained as
\begin{equation}
u = 1/l + (e/l)cos\Theta + k l^2 \Theta/h^3 + kl^2 e\Theta cos\Theta /h^3\label{e16}
\end{equation}
Keeping terms up to the power of '$e$' and $(k/\mu_0 )^2$, the time period '$\tau$' for one revolution is given to this order of approximation by
\begin{equation}
\tau = 2\pi L^2/h\label{e17}
\end{equation}
From (\ref{e16})
\begin{equation}
L  = l - \frac{kl^4\Theta}{h^3}\label{e18}
\end{equation}
Substituting in (\ref{e17}) we have
\begin{equation}
\tau = \frac{2\pi}{h} \left(l^2 - \frac{2kl^5\Theta}{h^3}\right)\label{e19}
\end{equation}
The second term in (\ref{e19}) represents the change in time period for one revolution. The decrease of time period is given by
\begin{equation}
\delta \tau = 8\pi^2 l^3 k/\mu_0^2\label{e20}
\end{equation}
The second term in (\ref{e18}) indicates the decrease in latus-rectum.\\
For one revolution the change of latus-rectum is given by
\begin{equation}
\delta l = 2\pi kl^{2.5}/\mu^{1.5}_0\label{e21}
\end{equation}
In the solar system, we have,
$$k = 8.988 mks units/sec.$$
Using $k$ and $\mu_0$ to find the change in time period and the latus rectum in the varying $G$ case by substituting in (\ref{e20}) and (\ref{e21}) respectively for Mercury we get
$$\delta T = 1.37 \times 10^{-5} sec/rev$$  
\begin{equation}
\delta l = 4.54 cm/rev\label{e22}
\end{equation}
We observe that the equations (\ref{e20}), (\ref{e21}) or (\ref{e22}) show a decrease in distance and in the time of revolution. If we use for the planetary motion, the General Relativistic analogue of (\ref{e14}), viz., 
$$\frac{d^2 u}{d\Theta^2} + u = \frac{\mu_0}{h^2} (1 + 3h^2 u^2),$$
then while we recover the precession of the perihelion of Mercury, for example, there is no effect similar to (\ref{e20}), (\ref{e21}) or (\ref{e22}). On the other hand this effect is very minute and only protracted careful observations can detect it.\\
But as noted, the decrease of the period in (\ref{e20}) has been observed in the case of Binary Pulsars.
\section{The Riddle of Gravitation}
Let us now come back to the puzzle which gravitation poses. 
We had already argued from different points of view to arrive at the otherwise empirically known equations \cite{psp,psu,ijmpa}
$$R = \sqrt{N} l_P = \sqrt{\bar{N}N} l$$
\begin{equation}
l = \sqrt{n} l_P\label{eb1}
\end{equation}
where $l_P, l$ and $R$ are the Planck length, the pion Compton wavelength and the radius of the universe and $N, \bar {N}$ and $n$ are certain large numbers. Some of these are well known empirically for example $\bar {N} \sim 10^{80}$ being the number of elementary particles, which typically are taken to be pions in the literature, in the universe.\\
One way of arriving at the above relations is by considering a series of $N$ Planck mass oscillators which are created out of the Quantum Vaccuum. In this case (Cf. also ref.\cite{ng}) we have
\begin{equation}
r = \sqrt{N a^2}\label{eb2}
\end{equation}
In (\ref{eb2}) $a$ is the distance between the oscillators and $r$ is the extent. If in (\ref{eb2}), $r$ is taken as $R$ and $a$ is taken as $l_P$ and similarly $a$ is taken as $l$ and $N$ as $\bar{N}$, we get back (\ref{eb1}). The supposedly empirical equations (\ref{eb1}) follow from equation (\ref{eb2}).\\
There is another way of arriving at equations (\ref{eb1}) (Cf.ref.\cite{ijmpa}). For this, we observe that the position operator for the Klein-Gordan equation is given by \cite{sch},
$$\vec X_{op} = \vec x_{op} - \frac{\imath \hbar c^2}{2} \frac{\vec p}{E^2}$$
Whence we get
\begin{equation}
\hat X^2_{op} \equiv \frac{2m^3 c^4}{\hbar^2} X^2_{op} - \alpha = \frac{2m^3 c^6}{\hbar^2} x^2 + \frac{p^2}{2m}\label{eb3}
\end{equation}
where $\alpha$ is a constant, irrelevant scalar. It can be seen that purely mathematically (\ref{eb3}) for $\hat {X}^2_{op}$ defines the Harmonic oscillator equation, this time with quantized, what may be called space levels. It turns out that these levels are all multiples of $(\frac{\hbar}{mc})^2$. This Compton length is the Planck length for a Planck mass particle. Accordingly we have for any system of extension $r$,
$$r^2 \sim Nl^2$$
which gives back equation (\ref{eb1}). It is also known that the Planck length is also the Schwarzschild radius of a Planck mass, that is we have
\begin{equation}
l_P = 2 Gm_P/c^2\label{eb4}
\end{equation}
Using equations (\ref{eb1}) and (\ref{eb4}), we will now deduce a few new and valid and a number of otherwise empirically known relations involving the various microphysical parameters and large scale parameters. Some of these relations are deducible from the others. Many of these relations featured (empirically) in Dirac's Large Number Cosmology. We follow Dirac and Melnikov in considering $l, m, \hbar, l_P, m_P$ and $e$ as microphysical parameters \cite{narlikar,melnikov}. Large scale parameters include the radius and the mass of the universe, the number of elementary particles in the universe and so on.\\
In the process we will also examine the nature of gravitation. It must also be observed that the Large Number relations below are to be considered in the Dirac sense, wherein for example the difference between the electron and pion (or proton) masses is irrelevant \cite{weinberg}.\\
We will use the following well known equation which has been obtained through several routes (Cf. for example \cite{hayakawa,nottale,ruffini}):
\begin{equation}
2 \frac{GM}{c^2} = R\label{eb5}
\end{equation}
We now observe that from the first two relations of (\ref{eb1}), using the Compton wavelength expression we get
\begin{equation}
m = m_P/\sqrt{n}\label{eb6}
\end{equation}
Using also the second relation in (\ref{eb1}) we can easily deduce
\begin{equation}
N = \bar {N} n\label{eb7}
\end{equation}
Using (\ref{eb1}) and (\ref{eb5}) we have
\begin{equation}
M = \sqrt{N} m_P\label{eb8}
\end{equation}
Interestingly (\ref{eb8}) can be obtained directly, without recourse to (\ref{eb5}), from the energy of the Planck oscillators (Cf.ref.\cite{psu}). Combining (\ref{eb8}) and (\ref{eb6}) we get
\begin{equation}
M = \left(\sqrt{N n}\right) m\label{eb9}
\end{equation}
Further if we use in the last of equation (\ref{eb1}) the fact that $l_P$ is the Schwarzchild radius that is equation (\ref{eb4}), we get, 
\begin{equation}
G = \frac{lc^2}{nm}\label{eb10}
\end{equation}
We now observe that if we consider the gravitational energy of the $N$ Planck masses (which do not have any other interactions) we get,
$$\mbox{Gravitational \, Energy}\, = \frac{GNm^2_P}{R}$$
If this is equated to the inertial energy in the universe, $Mc^2$, as can be easily verified we get back (\ref{eb5}). In other words the inertial energy content of the universe equals the gravitational energy of all the $N$ Planck oscillators.\\
Similarly if we equate the gravitational energy of the $n$ Planck oscillators constituting the pion we get
\begin{equation}
\frac{Gm^2_Pn}{R} = mc^2\label{eb11}
\end{equation}
Using in (\ref{eb11}) equation (\ref{eb4}) we get
$$\frac{l_Pm_Pn}{R} = m$$
Whence it follows on using (\ref{eb7}), (\ref{eb6}) and (\ref{eb1}),
\begin{equation}
n^{3/2} = \sqrt{N}, \, n = \sqrt{\bar {N}}\label{eb12}
\end{equation}
Substituting the value for $n$ from (\ref{eb12}) into (\ref{eb10}) we will get
\begin{equation}
G = \frac{lc^2}{\sqrt{\bar {N}}m}\label{eb13}
\end{equation}
Using in (\ref{eb13}), the expression for the Compton length,
$$l = \hbar / mc$$
and further, the fact that $\hbar c \sim e^2$, we get
\begin{equation}
Gm^2  = \frac{e^2}{\sqrt{\bar{N}}}\label{eb15}
\end{equation}
Equation (\ref{eb15}) is another empirically well known equation which was used by Dirac in his Cosmology.\\
Interestingly also rewriting (\ref{eb13}) as
$$G = \frac{l^2c^2}{Rm}$$
wherein we have used (\ref{eb1}) and further using the fact that $H = c/R$, where $H$ is the Hubble constant we can deduce
\begin{equation}
m \approx \left(\frac{H\hbar^2}{Gc}\right)^{\frac{1}{3}}\label{eb16}
\end{equation}
Equation (\ref{eb16}) is the so called mysterious Weinberg formula, known empirically \cite{weinberg}. As Weinberg put it, ``...it should be noted that the particular combination of $\hbar , H, G$, and $c$ appearing (in the formula) is very much closer to a typical elementary particle mass than other random combinations of these quantities; for instance, from $\hbar , G$, and $c$ alone one can form a single quantity $(\hbar c/G)^{1/2}$ with the dimensions of a mass, but this has the value $1.22 \times 10^{22} MeV/c^2$, more than a typical particle mass by about $20$ orders of magnitude!\\
``In considering the possible interpretations (of the formula), one should be careful to distinguish it from other numerical ``coincidences''... In contrast, (the formula) relates a single cosmological parameter, $H$, to the fundamental constants $\hbar , G, c$ and $m$, and is so far unexplained.''\\
We will come back to this point but remark that (\ref{eb13}) brings out gravitation in a different light-- somewhat on the lines of Sakharov. In fact it shows up gravitation as the excess or residual energy in the universe.\\
Finally it may be observed that (\ref{eb13}) can also be rewritten as 
\begin{equation}
\bar {N} = \left(\frac{c^2 l}{mG}\right)^2 \sim 10^{80}\label{ebx}
\end{equation}
and so also (\ref{eb10}) can be rewritten as
$$n = \left(\frac{lc^2}{Gm}\right) \sim 10^{40}$$
It now immediately follows that
$$N \sim 10^{120}$$
a result that also follows from (\ref{eb1}) itself (Cf. also \cite{ijmpe}). Looking at it this way, given $G$ and the microphysical parameters we can deduce the numbers $N, \bar{N}$ and $n$.\\
Thus the many so called large number coincidences and the mysterious Weinberg formula can be deduced on the basis of a Planck scale underpinning for the elementary particles and the whole universe. Moreover this also explains the Weinberg puzzle in (\ref{eb16}) - as can be seen from (\ref{eb13}), $G$ itself contains a large scale parameter viz., the number of particles in the universe so that $H$ is not the only cosmological parameter in (\ref{eb16}). This was done from a completely different point of view, namely using fuzzy spacetime and fluctuations in the 1997 model that as pointed out successfully predicted a dark energy driven accelerating universe with a small cosmological constant \cite{ijmpa,cu}.\\
However the above treatment brings out the role of the Planck scale oscillators in the Quantum Vaccuum. It resembles, as remarked earlier the Sakharov-Zeldovich metric elasticity of space approach \cite{sakharov}. Essentially Sakharov argues that the renormalization process in Quantum Field Theory which removes the Zero Point energies is altered in General Relativity due to the curvature of spacetime, that is the renormalization or subtraction no longer gives zero but rather there is a residual energy similar to the modification in the molecular bonding energy due to deformation of the solids. We see this in a little more detail following Wheeler \cite{mwt}. The contribution to the Lagrangian of the Zero Point energies can be given in a power series as follows
$${\it L} (r) = A\hbar \int k^3 dk + B\hbar^{(4)} r \int k dk$$
$$+ \hbar [C(^{(4)} r)^2 + Dr^{\alpha \beta} r_{\alpha \beta}] \int k^{-1} dk$$
\begin{equation}
+ (\mbox{higher-order \, terms}).\label{eb17}
\end{equation}
where $A, B, C$ etc. are of the order of unity and $r$ denotes the curvature. By renormalization the first term in (\ref{eb17}) is eliminated. According to Sakharov, the second term is the action principle term, with the exception of some multiplicated factors. (The higher terms in (\ref{eb17}) lead to corrections in Einstein's equations). Finally Sakharov gets
\begin{equation}
G = \frac{c^3}{16\pi B \hbar \int k dk}\label{eb18}
\end{equation}
Sakharov then takes a Planck scale cut off for the divergent integral in the denominator of (\ref{eb18}). This immediately yields
\begin{equation}
G \approx \frac{c^3 l^2_P}{\hbar}\label{eb19}
\end{equation}
Infact using relations like (\ref{eb1}), (\ref{e6}) and (\ref{eb12}), it is easy to verify that (\ref{eb19}) gives us back (\ref{eb10}) (and (\ref{eb13})).\\
According to Sakharov (and (\ref{eb19})), the value of $G$ is governed by the Physics of Fields and Particles and is a measure of the metrical elasticity at small spacetime intervals. It is a microphysical constant.\\
However in our interpretation of (\ref{eb13}) (which is essentially the same as Sakharov's equation (\ref{eb19})), $G$ appears as the expression of a residual energy over the entire universe: The entire universe has an underpinning of the $N$ Planck oscillators and is made up of $\bar {N}$ elementary particles, which again each have an underpinning of $n$ Planck oscillators. It must be reiterated that (\ref{eb19}) obtained from Sakharov's analysis shows up $G$ as a microphysical parameter because it is expressed in their terms. This is also the case in Dirac's cosmology.  This is also true of (\ref{eb10}) because $n$ relates to the micro particles exclusively.\\
However when we use the relation (\ref{eb12}), which gives $n$ in terms of $\bar {N}$, that is links up the microphysical domain to the large scale domain, then we get (\ref{eb13}). With Sakharov's equation (\ref{eb19}), the mysterious nature of the Weinberg formula remains. But once we use (\ref{eb13}), we are effectively using the large scale character of $G$ -- it is not a microphysical parameter. This is brought out by (\ref{ebx}), which is another form of (\ref{eb13}). If $G$ were a microphysical parameter, then the number of elementary particles in the universe would depend solely on the microphysical parameters and would not be a large scale parameter. The important point is that $G$ relates to elementary particles and the whole universe \cite{bgs}. That is why (\ref{eb13}) or equivalently the Weinberg formula (\ref{eb16}) relate supposedly microphysical parameters to a cosmological parameter. Once the character of $G$ as brought out by (\ref{eb13}) is recognized, the mystery disappears.\\
Finally it may be remarked that attempts to unite gravitation with other interactions have been unsuccessful for several decades. However, it is possible to get a description of gravitation in an extended gauge field formulation using noncommutative geometry (to take account of the fact that the graviton is a spin 2 particle) \cite{bgsannales,uof}.
\section{Discussion}
1. With regard to the time variation of $G$, it must be mentioned that without reference to the tests alluded to, different observations have yielded different values. Observations on the earth, in the solar system and with Pulsars have yielded for $\frac{\dot{G}}{G}$ a value $\sim 10^{-10}/yr$ as in (\ref{e2}). However other model dependent observations have yielded values $\sim 10^{-11}$ and $10^{-12}$ \cite{uzan}.\\
2. It may also be reiterated that other major effects like bending of light and the precession of the perihelion of Mercury have been shown to follow from (\ref{e2}) by using similar considerations \cite{nc115}. Moreover we can also show that it is possible to bypass dark matter in explaining the gravitational rotation curves \cite{cu}.\\
3. Another interesting consequence of the time variation of $G$ given in (\ref{e2}) is that it can be shown that this leads to an immediate transition from the Planck scale to the Compton scale by invoking considerations from black hole thermodynamics \cite{bhtd}.\\
4. Apart from the well known coincidences, it has recently been pointed out that there is a new coincidence \cite{xmpla} viz., the fact that there is a residual energy $10^{-33}eV$ which equals the Hubble radius (We are using natural units). This new puzzle is easily explained by the fact that there is a minimum mass in the universe, using the Planck scale underpinning considerations - this is $10^{-65}gms$. This also turns out to be the minimum thermodynamic mass in the universe (Cf.\cite{uof}), and has been identified with the mass of a photon \cite{bhtd,mp}. One can immediately identify the above energy of $10^{-33}eV$ with the energy of this minimum mass, and as can be easily verified the Hubble radius is the Compton length of this mass. Thus the above ``coincidence'' is symptomatic of the minimum mass or energy in the universe with its corresponding extent.   

\end{document}